\documentclass[runningheads]{llncs}

\usepackage[T1]{fontenc}
\usepackage{graphicx}
\usepackage{hyperref}
\usepackage{color}

\usepackage{amssymb}

\begin{document}

\title{Recent Developments in Real Quantifier Elimination and Cylindrical Algebraic Decomposition \\ (Extended Abstract of Invited Talk) }
\titlerunning{Recent Developments in Real QE and CAD (Invited Talk)}

\author{Matthew England \orcidID{0000-0001-5729-3420}}
\authorrunning{M. England}

\institute{Coventry University, Coventry, UK
\email{Matthew.England@coventry.ac.uk}\\
\url{https://matthewengland.coventry.domains} 
}

\maketitle              

\begin{abstract}
This extended abstract accompanies an invited talk at CASC 2024, which surveys recent developments in Real Quantifier Elimination (QE) and Cylindrical Algebraic Decomposition (CAD).  After introducing these concepts we will first consider adaptations of CAD inspired by computational logic, in particular the algorithms which underpin modern SAT solvers.  CAD theory has found use in collaboration with these via the Satisfiability Modulo Theory (SMT) paradigm; while the ideas behind SAT/SMT have led to new algorithms for Real QE.  
Second we will consider the optimisation of CAD through the use of Machine Learning (ML).  The choice of CAD variable ordering has become a key case study for the use of ML to tune algorithms in computer algebra. We will also consider how explainable AI techniques might give insight for improved computer algebra software without any reliance on ML in the final code.

\keywords{Quantifier Elimination \and Cylindrical Algebraic Decomposition \and SAT \and SMT \and Machine Learning \and Explainable AI.}
\end{abstract}

\section{Real Quantifier Elimination}
\label{sec:QE}

Quantifier Elimination (QE) may be considered as a form of simplification in mathematical logic:  given a quantified logical statement QE will produce a statement which is equivalent and does not involve any logical quantifiers (there exists / for all statements).  Real QE refers to the case where the logical atoms are constraints on polynomials over the real numbers.

For example, QE would convert the quantified statement $\exists x, x^2 + 3x + 1 > 0$ to \texttt{True} (consider e.g. $x=0$); while the quantified statement $\forall x, x^2 + 3x + 1 > 0$ is converted to \texttt{False} (consider e.g. $x=-1$).  
What about when there are unquantified variables in the formula?  Consider for example  $\forall x, x^2 + bx + 1 > 0$.  We just saw this was \texttt{False} when $b=3$ but we can also see it is \texttt{True} when $b=0$.  The truth depends on the unquantified $b$.  Using a Real QE procedure we may uncover that an equivalent unquantified statement is $-2 < b < 2$.  I.e. the original statement is \texttt{True} for such values of $b$ and \texttt{False} otherwise.

In the 1940s Tarski demonstrated that Real QE is always possible \cite{Tarski1948}, but it took decades before an algorithm could achieve Real QE in practice.

\section{Cylindrical Algebraic Decomposition}
\label{sec:CAD}

Collins introduced Cylindrical Algebraic Decomposition (CAD) as a method to solve the Real QE problem in the 1970s \cite{Collins1975}.  A CAD is a decomposition of $\mathbb{R}^n$ into connected subsets known as cells.  Each cell is a semi-algebraic set and the cells are arranged cylindrically, meaning the projection of any pair of such cells (with respect to the declared variable ordering) is either equal or disjoint.  I.e. the cells stack up in cylinders over cells in $\mathbb{R}^{n-1}$.  Collins' CAD is produced relative to a set of input polynomials and guarantees that each of these polynomials has invariant sign (negative, zero, or positive) throughout any such cell.   
The sign invariant decomposition allows us to draw conclusions over an infinite space by observing behaviour at a finite number of sample points (one per cell);  the cylindricity allows us to easily project and check membership of cells, and the semi-algebraic property means we can construct solution formulae from them.

\begin{figure}[t]
    \centering
    \includegraphics[width=0.48\textwidth]{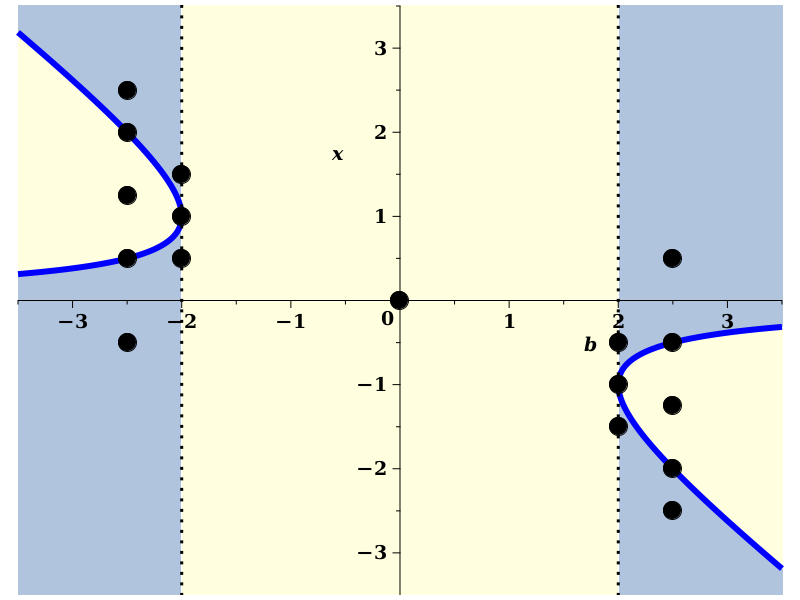}
    \includegraphics[width=0.48\textwidth]{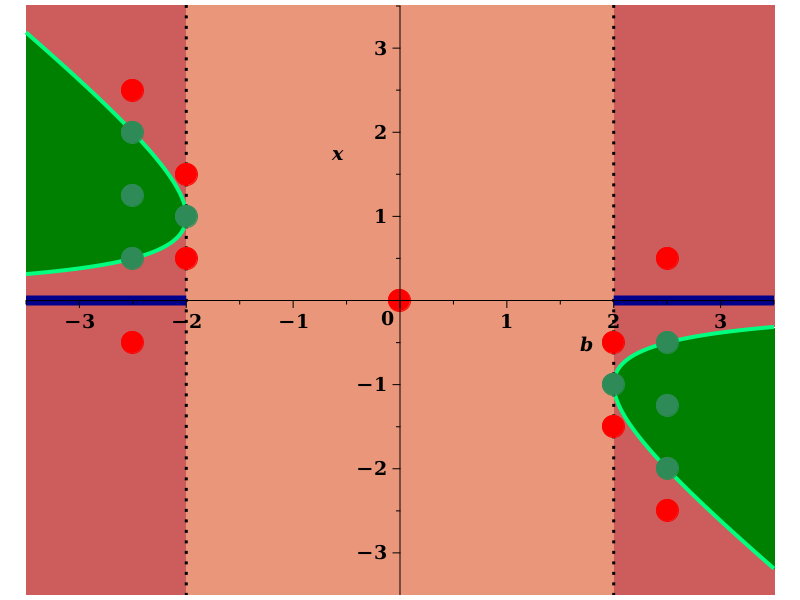}
    \caption{Visualisation of sign-invariant CADs built for the polynomial $x^2+bx+1$.  
    \\
    On the left: the polynomial is graphed as the dark solid curve with the seven 2D cells in the CAD the regions separated by the graph and the additional dotted lines.  Note that the CAD also contains two point cells (the points where the tangent of the polynomial is vertical); and eight 1D cells (the curve segments of the polynomial's graph either side of the point cells and the vertical line segments at $x = \pm 2$ either side of the point cells). Each cell contains a sample point (solid circles).   
    On the right:  the sample points have been shaded green (lighter in greyscale) and red (darker in greyscale)  depending on the truth of $x^2 + bx + 1 \leq 0$.  The true cells have been projected onto the $b$-axis.}
    \label{fig:QEviaCAD}
\end{figure}

An existential QE problem may be tackled by building a CAD for the polynomials involved, testing to identify cells where the formulae is satisfied, projecting true cells, and then taking their union.  For example, consider $\exists x, x^2 + bx + 1 \leq 0$.  A sign-invariant CAD for the polynomial in the formula is shown on the left of Figure \ref{fig:QEviaCAD}.  On the right the cells have been shaded according to the truth of the formula and projecting the true cells identified $b \leq -2 \lor b \geq 2$.   These are the values of our unquantified variable for which there exists a suitable $x$.

\noindent A universal QE problem is tackled by using the transformation
\[
\forall x P(x) = \lnot \exists x \lnot P(x)
\]
to convert it to an existential QE problem whose answer we then negate.  Recall our original example from Section \ref{sec:QE} was $\forall x, x^2 + bx + 1 > 0$.  The translation leads us to study $\exists x, x^2 + bx + 1 \leq 0$ whose solution we just uncovered.  Negating that gives $-2 < b \land b<+2$, as previously claimed in Section \ref{sec:QE}.   The CAD structure is well suited for QE as we can project and take complements of cells with ease.

\section{The Doubly Exponential Wall}

\begin{figure}[h]
    \centering
    \includegraphics[width=0.99\textwidth]{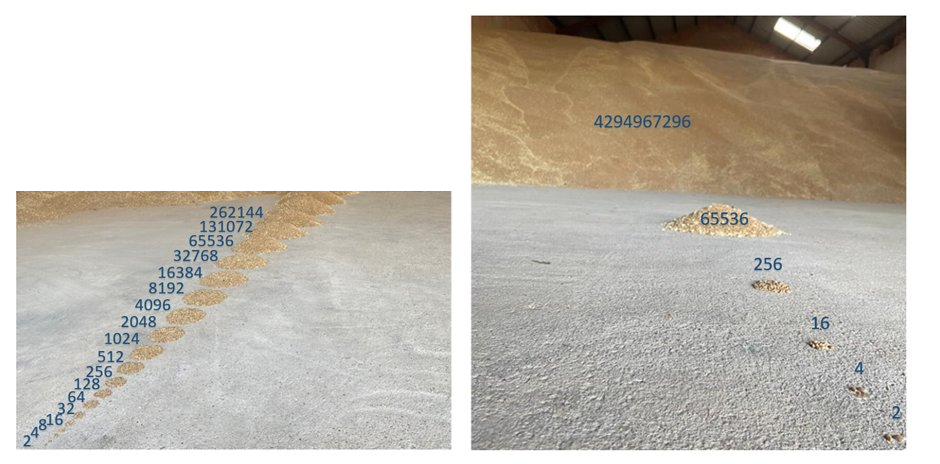}
    \caption{These piles of grain demonstrate exponential growth (left) and doubly exponential growth (right).  Image credit:  Tereso del R\'{i}o.}
    \label{fig:doubleE}
\end{figure}

CAD is probably the best known general purpose method for Real QE.  However, CAD has doubly exponential complexity \cite{BD07}, in effect producing a wall beyond which its application is infeasible.  This is visualised in comparison to exponential growth in Figure \ref{fig:doubleE}.  In the almost half century since its inception there has been a tremendous amount of research to improve the performance of CAD.  The first 20 years of progress were summarised in the book \cite{CJ98} and in particular the article \cite{Collins1998a}.  For work in the subsequent decades see for example the introduction of \cite{BDEMW16}.  None of these improvements have addressed the fundamental complexity of CAD; however, they have brought a great many more applications into scope of CAD, in effect "\emph{pushing back}" that doubly exponential wall.  

In this talk we will consider two recent developments in CAD theory which push the doubly exponential wall further still.  Both integrate of CAD with other sub-fields of Computer Science and we hypothesise that other areas of symbolic computation and computer algebra may benefit from similar integrations.

\section{CAD and Satisfiability Checking}

Consider a restriction of the QE problem to the case where all variables are existentially quantified, i.e. to identify whether
\[
\exists x_1, \exists x_2, \dots, \exists x_n F(x_1, x_2, \dots, x_n)
\]
(for some logical formula $F$) is either \texttt{True} (satisfiable, SAT) or \texttt{False} (unsatisfiable, UNSAT).  This is the Satisfiability Checking problem.  In the case where the atoms of the formulae $F$ are multivariate polynomials constraints then this may be solved using a CAD for the polynomials involved.  How can we adapt CAD to benefit from the simpler logical structure?

\subsection{SAT and SMT}

In the case where the atoms of $F$ are Boolean then we have the original SAT problem.  Although famously NP-complete, there now exist SAT-solvers which can routinely tackle extremely large SAT problems in practice.  
The success of SAT solvers stems from their sophisticated search algorithms to process the exponential search space, see for example \cite{BHvMW09}.  Such is the success, that they are now used to tackle long-standing open problems in mathematics \cite{HK17a}.  

\begin{figure}[h]
    \centering
    \includegraphics[width=0.7\textwidth]{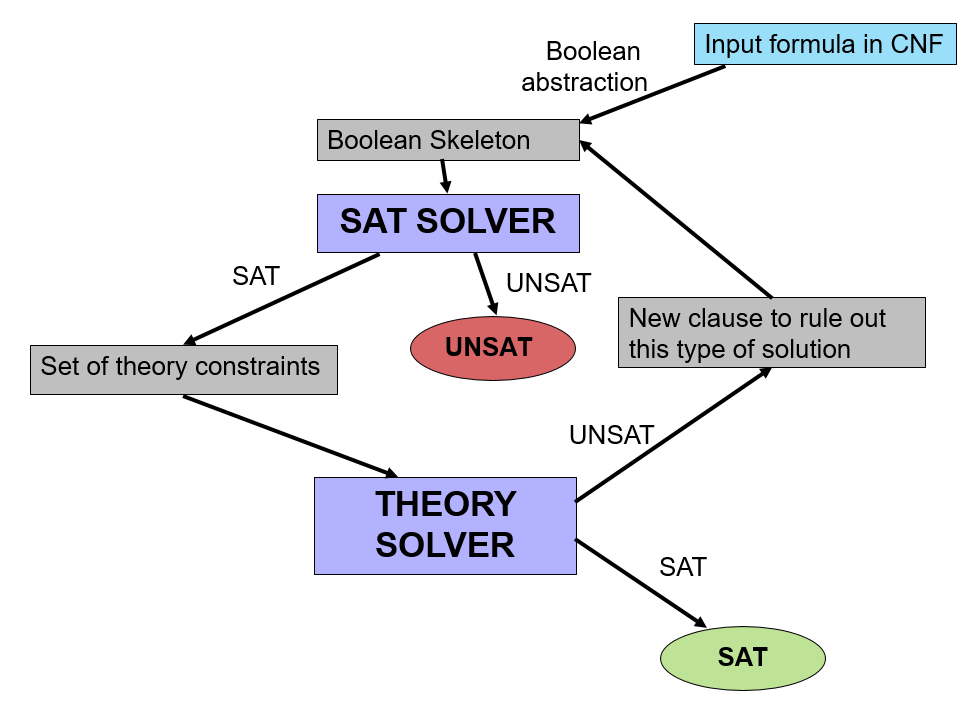}
    \caption{Schematic of the Satisfiability Modulo Theory Paradigm}
    \label{fig:SMT}
\end{figure}

The Satisfiability Modulo Theory (SMT) paradigm seeks to apply SAT-solvers on more general problems by combining them with a theorem solver which can check the compatibility of a set of atoms in the domain of interest.  They work in a loop as visualised in Figure \ref{fig:SMT} with the SAT-solver proposing solutions to the Boolean skeleton of the formula to be checked by the theory solver.  If found unsuitable the skeleton is extended to rule out such solutions.  

\subsection{CAD as Theory Solver}

CAD is one possibility for theory solver in the case of SMT for non-linear real arithmetic.  However, for this to be efficient the CAD implementation should support incremental calls, backtracking and minimal explanation generation when there is no solution \cite{KA20}.  For problems where the solution is SAT this approach can determine the solution much faster than CAD alone as an entire decomposition need be constructed.  For UNSAT problems this approach can still give benefits by allowing us to reach the conclusion through the study of multiple smaller problems rather than one big problem:  beneficial for an algorithm with double exponential complexity.

\subsection{New CAD-based algorithms}

\begin{figure}
    \centering
    \includegraphics[width=0.95\textwidth]{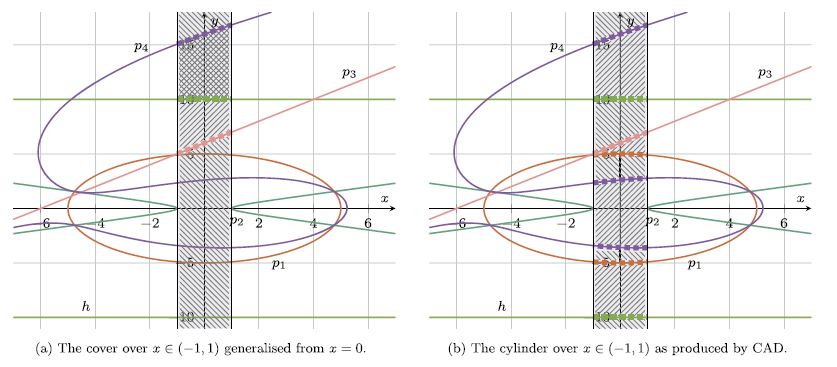}
    \caption{Figure 15 from \cite{ADEK21}.  Demonstrating that a fewer number of polynomials (only those whose intersection with the shaded region is dotted) may be used to produce a covering than a decomposition (which would use all polynomials graphed).}
    \label{fig:covering}
\end{figure}

Subsequently, the present author and colleagues redesigned the CAD based SMT theory solver to make use of the key ideas behind SAT/SMT in \cite{ADEK21}. Like a SAT-solver this takes a search-based approach: choosing a sample point not yet considered and generalising the findings at that point to a cell around the point using CAD theory.  The cells produced gradually form of covering of $\mathbb{R}^n$, rather than a decomposition, which can be achieved with a smaller number of larger cells, requiring less computation as visualised in Figure \ref{fig:covering}.  Subsequent sample points are chosen from outside cells already generated, in effect guiding the search away from unproductive areas of the search space, analogous to the \emph{conflict driven clause learning} inside SAT solvers.  This cylindrical algebraic covering concept has since been extended to tackle the original Real QE problem from Section \ref{sec:QE} in \cite{KN23}.  

\subsection{SC$^2$}

This is just one example of the integration of computer algebra systems and SAT/SMT solvers as supported by the SC$^2$ initiative \cite{AAB+16a}. SAT-solvers have also inspired the NuCAD method for Real QE \cite{Brown2015}, while the new MCSAT proof framework for satisfiabilty checking (an alternative to the SMT paradigm) was originally developed with an adaption of CAD in \cite{JdM12}.  

Another recent influence of SAT/SMT to Real QE is the proof system presentation of algorithms, which allows for the elegant separation of mathematical correctness from heuristic choices to be optimised \cite{NASBDE24}.

\section{CAD and Machine Learning}

Machine Learning (ML) uses statistics upon large quantities of data to learn how to perform tasks that have not been explicitly programmed.  ML technology is behind most recent Artificial Intelligence (AI) applications.  It is natural to ask whether ML can help with symbolic computation?

An example of ML to perform tasks traditionally undertaken by computer algebra is \cite{LC20} which used a transformer to perform function integration and solve ODEs.  However, those tasks are cheap to symbolically check a proposed answer, allowing us to mitigate any risks from the ML being wrong:  this is not case for most computer algebra tasks, including Real QE.  For most tasks we suggest it may be better to use computer algebra in tandem with ML, by allowing the ML to tune or guide the symbolic computation algorithm.  Some of the challenges to overcome here include how best to \emph{embed} the algebra problem for the ML tools; how to minimise the amount of costly symbolic computation labelling of data; and how to generate representative data sets?

\subsection{CAD Variable Ordering Choice}

\begin{figure}[ht]
    \centering
    \includegraphics[width=0.45\textwidth]{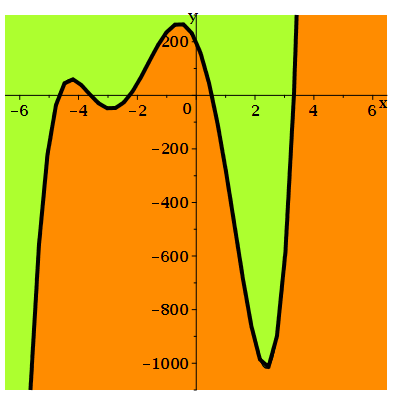}
    \includegraphics[width=0.48\textwidth]{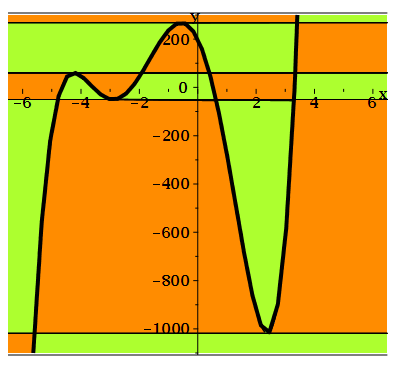}
    \caption{Two CADs for the same polynomial with full dimensional cells the connected regions between the graph and the additional horizontal lines.  On the left the cells project into a single cylinder over the whole horizontal axis; while on the right they project onto multiple cylinders over the vertical axis. }
    \label{fig:CADord}
\end{figure}

One choice in the CAD algorithm that can benefit from tuning is the variable ordering.  This determines the order of processing by the algorithm (and is used in the definition of cylindricity).  Depending on the application there is often freedom in the choice:  for Real QE we must order the variables as they are quantified but have freedom inside quantifier blocks (and in the order of the unquantified variables).  The variable ordering can make a substantial difference to the number of cells produced as visualised in Figure \ref{fig:CADord}, and can effect the tractibiloty and even the complexity of CAD \cite{BD07}.

The first paper to apply ML to tune any sort of computer algebra was in fact a 2014 study to select the variable ordering for CAD \cite{HEWDPB14} (in fact, to select which of three human-designed heuristics to follow for the choice).  In the decade that followed the CAD variable ordering choice has become something of a case study for ML optimisation of computer algebra with experiments since including different models \cite{EF19}, different embedding \cite{FE19}, deep learning \cite{CZC20} and reinforcement learning \cite{JDLHMZ23}.  We note recent lessons on the need for care over the dataset and the benefits of data augmentation \cite{dRE23}.

\subsection{Future Progress from Explainable AI?}

Understandably, some computer algebra developers would prefer to avoid the use of ML in their systems, to avoid code/data dependencies, to ensure deterministic behaviour, and to avoid reliance on a black box.  We finish by pointing out that such developers may still gain insight from the results of ML. 

In \cite{PSH20} the authors applied reinforcement learning to choose the order in which to process $S$-pairs in Buchberger's algorithm for a Gr\"{o}bner Basis.  Their analysis in \cite[{\S}5.1]{PSH20}  of the ML model revealed some simple strategies that accounted for much of the model's benefit.  Although simple, these strategies were not previously documented in the literature.  Developers who read this could implement and use these strategies without any use of ML themselves.

This led to our recent work in \cite{PdREC24} which sought to use Explainable AI (XAI) tools to automate such analysis for the case of CAD variable ordering choice.  We applied the SHAP tool \cite{LL17} to explain our previous work and had this identify the most important features in decision making.  Using a small number of these in a simple decision tree heuristic which may be implemented in a few lines of code, we outperform the previous (non-ML) based state-of-the-art heuristic for the CAD variable ordering choice.  

We hypothesise that this approach $-$ the development of human-designed heuristics following suggestions from the use of XAI tools to explain an ML choice for the decision $-$ may be a profitable new methodology for heuristic design within computer algebra systems.  

\newpage

\subsection*{Acknowledgments}

The author thanks the coauthors of the surveyed works:  Erika \'{A}brah\'{a}m, Kelley Cohen, James H. Davenport, Tereso del R\'{i}o, Dorian Florescu, Gereon Kremer, Jasper Nalbach, and Lynn Pickering.  

The author is currently supported by UKRI EPSRC grant EP/T015748/1: \emph{Pushing Back the Doubly-Exponential Wall of Cylindrical Algebraic Decomposition} (the DEWCAD Project).  Some of the work surveyed was supported by UKRI EPSRC grant EP/R019622/1, \emph{Embedding Machine Learning within Quantifier Elimination Procedures}; and EU H2020 FET CSA 712689, \emph{Satisfiability Checking and Symbolic Computation} (the SC$^2$ Project).

The author thanks the organisers of CASC 2024 for the invitation to speak.

\bibliographystyle{splncs04}
\bibliography{CAD}

\begin{thebibliography}{10}
\providecommand{\url}[1]{\texttt{#1}}
\providecommand{\urlprefix}{URL }
\providecommand{\doi}[1]{https://doi.org/#1}

\bibitem{AAB+16a}
{\'A}brah{\'a}m, E., Abbott, J., Becker, B., Bigatti, A.M., Brain, M.,
  Buchberger, B., Cimatti, A., Davenport, J.H., England, M., Fontaine, P.,
  Forrest, S., Griggio, A., Kroening, D., Seiler, W.M., Sturm, T.:
  $\mathsf{SC}^2$: Satisfiability checking meets symbolic computation. In:
  Kohlhase, M., Johansson, M., Miller, B., {de~Moura}, L., Tompa, F. (eds.)
  Intelligent Computer Mathematics: Proceedings CICM 2016, Lecture Notes in
  Computer Science, vol.~9791, pp. 28--43. Springer International Publishing
  (2016), \url{https://doi.org/10.1007/978-3-319-42547-4_3}

\bibitem{ADEK21}
{\'A}brah{\'a}m, E., Davenport, J.H., England, M., Kremer, G.: Deciding the
  consistency of non-linear real arithmetic constraints with a conflict driven
  search using cylindrical algebraic coverings. Journal of Logical and
  Algebraic Methods in Programming  \textbf{119},  100633 (2021),
  \url{https://doi.org/10.1016/j.jlamp.2020.100633}

\bibitem{BHvMW09}
Biere, A., Heule, M., {van Maaren}, H., Walsh, T.: Handbook of Satisfiability
  (Volume 185 Frontiers in Artificial Intelligence and Applications). IOS Press
  (2009)

\bibitem{BDEMW16}
Bradford, R., Davenport, J.H., England, M., McCallum, S., Wilson, D.: Truth
  table invariant cylindrical algebraic decomposition. Journal of Symbolic
  Computation  \textbf{76},  1--35 (2016),
  \url{http://dx.doi.org/10.1016/j.jsc.2015.11.002}

\bibitem{Brown2015}
Brown, C.W.: Open non-uniform cylindrical algebraic decompositions. In:
  Proceedings of the 2015 International Symposium on Symbolic and Algebraic
  Computation. pp. 85--92. ISSAC '15, ACM (2015),
  \url{https://doi.org/10.1145/2755996.2756654}

\bibitem{BD07}
Brown, C.W., Davenport, J.H.: The complexity of quantifier elimination and
  cylindrical algebraic decomposition. In: Proceedings of the 2007
  {I}nternational {S}ymposium on {S}ymbolic and {A}lgebraic {C}omputation. pp.
  54--60. ISSAC '07, ACM (2007), \url{https://doi.org/10.1145/1277548.1277557}

\bibitem{CJ98}
Caviness, B., Johnson, J.: Quantifier Elimination and Cylindrical Algebraic
  Decomposition. Texts \& Monographs in Symbolic Computation, Springer-Verlag
  (1998), \url{https://doi.org/10.1007/978-3-7091-9459-1}

\bibitem{CZC20}
Chen, C., Zhu, Z., Chi, H.: Variable ordering selection for cylindrical
  algebraic decomposition with artificial neural networks. In: Bigatti, A.,
  Carette, J., Davenport, J.H., Joswig, M., {de Wolff}, T. (eds.) Mathematical
  Software -- {ICMS} 2020. Lecture Notes in Computer Science, vol. 12097, pp.
  281--291. Springer International Publishing (2020),
  \url{https://doi.org/10.1007/978-3-030-52200-1_28}

\bibitem{Collins1975}
Collins, G.E.: Quantifier elimination for real closed fields by cylindrical
  algebraic decomposition. In: Proceedings of the 2nd GI Conference on Automata
  Theory and Formal Languages. pp. 134--183. Springer-Verlag (reprinted in the
  collection \cite{CJ98}) (1975),
  \url{https://doi.org/10.1007/3-540-07407-4_17}

\bibitem{Collins1998a}
Collins, G.E.: Quantifier elimination by cylindrical algebraic decomposition --
  20 years of progress. In: Caviness, B., Johnson, J. (eds.) Quantifier
  Elimination and Cylindrical Algebraic Decomposition, pp. 8--23. Texts \&
  Monographs in Symbolic Computation, Springer-Verlag (1998),
  \url{https://doi.org/10.1007/978-3-7091-9459-1_2}

\bibitem{dRE23}
{del R{\i}o}, T., England, M.: Data augmentation for mathematical objects. In:
  \'{A}brah\'{a}m, E., Sturm, T. (eds.) Proceedings of the 8th Workshop on
  Satisfiability Checking and Symbolic Computation ($\mathsf{SC}^2$ 2023). pp.
  29--38. No.~3455 in CEUR Workshop Proceedings (2023),
  \url{http://ceur-ws.org/Vol-3455/}

\bibitem{EF19}
England, M., Florescu, D.: Comparing machine learning models to choose the
  variable ordering for cylindrical algebraic decomposition. In: Kaliszyk, C.,
  Brady, E., Kohlhase, A., Sacerdoti, C.C. (eds.) Intelligent Computer
  Mathematics. Lecture Notes in Computer Science, vol. 11617, pp. 93--108.
  Springer International Publishing (2019),
  \url{https://doi.org/10.1007/978-3-030-23250-4_7}

\bibitem{FE19}
Florescu, D., England, M.: Algorithmically generating new algebraic features of
  polynomial systems for machine learning. In: Abbott, J., Griggio, A. (eds.)
  Proceedings of the 4th Workshop on Satisfiability Checking and Symbolic
  Computation ($\mathsf{SC}^2$ 2019). No.~2460 in CEUR Workshop Proceedings
  (2019), \url{http://ceur-ws.org/Vol-2460/}

\bibitem{HK17a}
Heule, M.J.H., Kullmann, O.: The science of brute force. Commun. ACM
  \textbf{60}(8),  70--79 (2017), \url{https://doi.org/10.1145/3107239}

\bibitem{HEWDPB14}
Huang, Z., England, M., Wilson, D., Davenport, J.H., Paulson, L., Bridge, J.:
  Applying machine learning to the problem of choosing a heuristic to select
  the variable ordering for cylindrical algebraic decomposition. In: Watt,
  S.M., Davenport, J.H., Sexton, A.P., Sojka, P., Urban, J. (eds.) Intelligent
  Computer Mathematics, Lecture Notes in Artificial Intelligence, vol.~8543,
  pp. 92--107. Springer International (2014),
  \url{http://dx.doi.org/10.1007/978-3-319-08434-3_8}

\bibitem{JDLHMZ23}
Jia, F., Dong, Y., Liu, M., Huang, P., Ma, F., Zhang, J.: Suggesting variable
  order for cylindrical algebraic decomposition via reinforcement learning. In:
  Thirty-seventh Conference on Neural Information Processing Systems (NIPS
  2023) (2023), \url{https://openreview.net/forum?id=vNsdFwjPtL}

\bibitem{JdM12}
Jovanovic, D., de~Moura, L.: Solving non-linear arithmetic. In: Gramlich, B.,
  Miller, D., Sattler, U. (eds.) Automated Reasoning: 6th International Joint
  Conference ({IJCAR}), Lecture Notes in Computer Science, vol.~7364, pp.
  339--354. Springer (2012), \url{https://doi.org/10.1007/978-3-642-31365-3_27}

\bibitem{KA20}
Kremer, G., {\'A}brah{\'a}m, E.: Fully incremental {CAD}. Journal of Symbolic
  Computation  \textbf{100},  11--37 (2020),
  \url{https://doi.org/10.1016/j.jsc.2019.07.018}

\bibitem{KN23}
Kremer, G., Nalbach, J.: Cylindrical algebraic coverings for quantifiers. In:
  Uncu, A., Barbosa, H. (eds.) Proceedings of the 7th Workshop on
  Satisfiability Checking and Symbolic Computation ($\mathsf{SC}^2$ 2022).
  pp.~1--9. No.~3458 in CEUR Workshop Proceedings (2023),
  \url{https://ceur-ws.org/Vol-3458}

\bibitem{LC20}
Lample, G., Charton, D.: Deep learning for symbolic mathematics. In: Mohamed,
  S., White, M., Cho, K., Song, D. (eds.) Eighth International Conference on
  Learning Representations (ICLR 2020) (2020),
  \url{https://iclr.cc/virtual_2020/poster_S1eZYeHFDS.html}

\bibitem{LL17}
Lundberg, S.M., Lee, S.I.: A unified approach to interpreting model
  predictions. In: Proceedings of the 31st International Conference on Neural
  Information Processing Systems (NIPS 2017). pp. 4768--4777. Curran Associates
  Inc. (2017), \url{https://dl.acm.org/doi/10.5555/3295222.3295230}

\bibitem{NASBDE24}
Nalbach, J., Abraham, E., Specht, P., Brown, C.W., Davenport, J.H., , England,
  M.: Levelwise construction of a single cylindrical algebraic cell. Journal of
  Symbolic Computation  \textbf{123},  102288 (2024),
  \url{https://doi.org/10.1016/j.jsc.2023.102288}

\bibitem{PSH20}
Peifer, D., Stillman, M., {Halpern-Leistner}, D.: Learning selection strategies
  in {B}uchberger's algorithm. In: {Daum\'{e} III}, H., Singh, A. (eds.)
  Proceedings of the 37th International Conference on Machine Learning (ICML
  2020). Proceedings of Machine Learning Research, vol.~119, pp. 7575--7585.
  PMLR (2020), \url{https://proceedings.mlr.press/v119/peifer20a.html}

\bibitem{PdREC24}
Pickering, L., {Del Rio Almajano}, T., England, M., Cohen, K.: Explainable {AI}
  insights for symbolic computation: {A} case study on selecting the variable
  ordering for cylindrical algebraic decomposition. Journal of Symbolic
  Computation  \textbf{123},  102276 (2024),
  \url{https://doi.org/10.1016/j.jsc.2023.102276}

\bibitem{Tarski1948}
Tarski, A.: A Decision Method For Elementary Algebra And Geometry. RAND
  Corporation, Santa Monica, CA (reprinted in the collection \cite{CJ98})
  (1948)

\end{thebibliography}

\end{document}